\documentclass[11pt]{article}
\usepackage[english]{babel}
\usepackage{verbatim}
\usepackage{authblk}
\usepackage{amsmath, amsthm, amssymb, graphics, graphicx}
\usepackage[shortlabels]{enumitem}
\usepackage{enumitem}

\newcommand{\mathsym}[1]{{}}

\newtheorem{noname}{}[section]

\title{\bf\Large{{Detecting induced subdivision of $K_4$}}}
\author{Ngoc Khang Le 
\thanks{This work was performed within the framework of the LABEX MILYON (ANR-10-LABX-0070) of Universit\'e de Lyon, within the program ``Investissements d'Avenir'' (ANR-11- IDEX-0007) operated by the French National Research Agency (ANR). Partially supported by ANR project Stint under reference ANR-13-BS02-0007.}
}
\affil{LIP, ENS de Lyon, Lyon, France}

\begin{document}
\maketitle

\begin{abstract}
In this paper, we propose a polynomial-time algorithm to test whether a given graph contains a subdivision of $K_4$ as an induced subgraph.
\end{abstract}

\section{Introduction}

We say that a graph $G$ \emph{contains} some graph $H$ if there exists an induced subgraph of $G$ isomorphic to $H$. A graph $G$ is \emph{$H$-free} if $G$ does not contain $H$. For $n\geq 1$, denote by $K_n$ the complete graph on $n$ vertices. A \emph{triangle} is a graph isomorphic to $K_3$. A \textit{subdivision} of a graph $G$ is obtained by subdividing its edges into paths of arbitrary length (at least one). A graph that does not contain any induced subdivision of $K_n$ is \textit{ISK$n$-free}. A \emph{twin wheel} is a graph consisting of a chordless cycle $C$ of length at least $4$ and a vertex with exactly three consecutive neighbors in $C$. Note that $K_4$ and twin wheels are two special kinds of ISK4. 

The class of ISK4-free graphs has recently been studied. In \cite{LMT12}, a decomposition theorem for this class is given. However it does not lead to a recognition algorithm. The chromatic number of this class is also proved to be bounded by $24$ in \cite{L16}, while it is conjectured to be bounded by $4$ \cite{LMT12}. Given a graph $H$, the \textit{line graph} of $H$ is the graph $L(H)$ with vertex set $E(G)$ and edge set $\{ef:e\cap f\neq \emptyset\}$. Since the class of ISK4-free graphs contains the line graph of every cubic graph, where finding the edge chromatic number is known to be NP-hard \cite{H81}, we know that finding the chromatic number of ISK4-free graphs is also NP-hard. 

For a fixed graph $H$, the question of detecting induced subdivision of $H$ in a given graph has been studied in \cite{LLMT09}. There are certain graphs $H$ where the problem is known to be NP-hard and graphs $H$ where there exists a polynomial algorithm. For example, detecting induced subdivision of $K_3$ is trivial since a graph is ISK3-free iff it is a forest. On the other hand, detecting induced subdivision of $K_5$ has been shown to be NP-hard \cite{LLMT09}. So far, apart from the trivial cases, the only two subcubic graphs that we have a polynomial-time algorithm to detect its induced subdivision are $K_{2,3}$ \cite{CS10} and net \cite{CST13}. Here, we answer the question of detecting ISK4, which was asked in \cite{LLMT09} and \cite{CST13}, by proving the following:

\begin{noname} \label{detect ISK4}
There is an algorithm with the following specifications:
\begin{itemize}
\item \emph{\bf Input}: Graph $G$. 
\item \emph{\bf Output}: 
	\begin{itemize}
		\item An ISK4 in $G$, or
		\item Conclude that $G$ is ISK4-free.
	\end{itemize} 
\item \emph{\bf Running time}: $O(n^9)$.
\end{itemize}
\end{noname}

\section{Preliminaries}

We first introduce some notions that we use in this paper. Let $G(V,E)$ be a graph with $n$ vertices. For $X\subseteq V(G)$, we denote by $G\setminus X$ the subgraph of $G$ induced by $V(G)\setminus X$. For $u\in V(G)$, let $N_G(u)$ denote the set of neighbors of $u$ in $G$ and $N_G[u]=N_G(u)\cup \{u\}$. We also extend that notion for a subset $X\subseteq V(G)$,  let $N_G(X)=\cup_{x\in X}N_G(x)\setminus X$ and $N_G[X]=N_G(x)\cup X$. If the context is clear, we write $N(u)$ and $N(X)$ instead of $N_G(u)$ and $N_G(X)$. For $k\geq 1$, a graph $P$ on $\{x_1,\ldots,x_k\}$ is a \textit{path} if $x_ix_j\in E(P)$ iff $|i-j|=1$ (this is often referred to an \emph{induced} or \emph{chordless path} in literature). The \textit{length} of a path is the number of its edges. The two \textit{ends} of $P$ are $x_1$ and $x_k$. The \textit{interior} of $P$ is $\{x_2,\ldots,x_{k-1}\}$. We denote by $x_iPx_j$ the subpath of $P$ from $x_i$ to $x_j$ and denote by $P^*$ the subpath of $P$ from $x_2$ to $x_{k-1}$ ($x_2Px_{k-1}$). A \emph{claw} is a graph on four vertices $\{u,x,y,z\}$ that has exactly three edges: $ux$, $uy$, $uz$. Vertex $u$ is called the \emph{center} of that claw. 

Let $x,y,z$ be three distinct pairwise non-adjacent vertices in $G$. A graph $H$ is an \emph{$(x,y,z)$-radar} in $G$ if it is an induced subgraph of $G$ and:
\begin{itemize}
	\item $V(H)=V(C)\cup V(P_x)\cup V(P_y)\cup V(P_z)$.
	\item $C$ is an induced cycle of length $\geq 3$ containing three distinct vertices $x'$, $y'$, $z'$.
	\item $P_x$ is a path from $x$ to $x'$, $P_y$ is a path from $y$ to $y'$, $P_z$ is a path from $z$ to $z'$.
	\item $P_x$, $P_y$, $P_z$ are vertex-disjoint and $x'$, $y'$, $z'$ are the only common vertices between them and $C$.
	\item These are the only edges in $H$.
\end{itemize}

Note that the length of each path $P_x$, $P_y$, $P_z$ could be $0$, therefore an induced cycle in $G$ passing through $x$, $y$, $z$ is considered as an $(x,y,z)$-radar. It is not hard to see that \ref{detect ISK4} is a direct consequence of the following algorithm.

\begin{noname} \label{detect radar}
There is an algorithm with the following specifications:
\begin{itemize}
\item \emph{\bf Input}: A graph $G$, four vertices $u,x,y,z\in V(G)$ such that $\{u,x,y,z\}$ induces a claw with center $u$ in $G$.
\item \emph{\bf Output}: One of the followings:
	\begin{itemize}
		\item An ISK4 in $G$, or
		\item Conclude that there is no $(x,y,z)$-radar in ${G'=G\setminus (N[u]\setminus \{x,y,z\})}$.
	\end{itemize}
\item \emph{\bf Running time}: $O(n^5)$.
\end{itemize}
\end{noname}

The following is trivial:

\begin{noname} \label{trivial lemma}
An ISK4 is either $K_4$, a twin wheel or contains a claw.
\end{noname}

\begin{proof}[Proof of \ref{detect ISK4} by \ref{detect radar}] 
We describe an algorithm to detect an ISK4 in $G$ as follows. First, we check if there is a $K_4$ or a twin wheel in $G$. Checking if there exists a $K_4$ takes $O(n^4)$. Checking if there is a twin wheel in $G$ can be done as follows: list all $4$-tuples $(a,b,c,d)$ of vertices in $G$ such that they induce a $K_4\setminus e$ (a graph obtained from $K_4$ by removing one edge, usually called a \emph{diamond}) where $ad\notin E(G)$; for each tuple, check if $a$ and $d$ are connected in $G\setminus ((N[b]\cup N[c])\setminus \{a,d\})$. Since we have $O(n^4)$ such tuples, this can be done in $O(n^6)$. If there exists a $K_4$ or a twin wheel in $G$, then output that ISK4 in $G$. Otherwise, move on to next step.

Now we may assume that $G$ is $\{K_4$, twin wheel$\}$-free. The following claim is true thanks to \ref{trivial lemma}: $G$ contains an ISK4 iff there exists some $4$-tuple $(u,x,y,z)$ of vertices in $G$ such that they induce a claw with center $u$ and there is an $(x,y,z)$-radar in $G'$. The last step in our algorithm is the following: generate every $4$-tuple $(u,x,y,z)$ of vertices in $G$ such that they induce a claw with center $u$ and run Algorithm \ref{detect radar} for each tuple. If for some tuple $(u,x,y,z)$, we detect an ISK4 in $G$ then output that ISK4 and stop. If for all the tuples, we conclude that there is no $(x,y,z)$-radar in $G'$ then we can conclude that $G$ contains no ISK4. Since we have $O(n^4)$ such tuples, and it takes $O(n^5)$ for each tuple by Algorithm \ref{detect radar} then the running time of our algorithm is $O(n^9)$. 
\end{proof}

The rest of our paper is therefore devoted to the proof of \ref{detect radar}. In the next section, we introduce some useful structures and the main proof is presented in Section \ref{S:4}.

\section{Antennas and cables} \label{S:3}

First we introduce two useful structures in our algorithm. 

Let $x$, $y$, $z$ be three distinct pairwise non-adjacent vertices in $G$. An \emph{$(x,y,z)$-antenna} in $G$ is an induced subgraph $H$ of $G$ such that:
\begin{itemize}
	\item $V(H)=\{c\}\cup V(P_x)\cup V(P_y)\cup V(P_z)$.
	\item $c\notin \{x,y,z\}\cup V(P_x)\cup V(P_y)\cup V(P_z)$.
	\item $P_x$ is a path from $x$ to $x'$, $P_y$ is a path from $y$ to $y'$, $P_z$ is a path from $z$ to $z'$.
	\item $P_x$, $P_y$, $P_z$ are vertex-disjoint and at least one of them has length $\geq 1$.
	\item $cx',cy',cz'\in E(H)$.
	\item These are the only edges in $H$.
	\item For any vertex $v$ in $G\setminus H$:
	\begin{itemize}
		\item $v$ has no neighbor in $H$ or exactly one neighbor in $H$, or
		\item $v$ has exactly two neighbors $v_1$, $v_2$ in $H$ such that for some $t\in\{x,y,z\}$, $v_1,v_2\in P_t\cup \{c\}$ and their distance in $H$ is $1$ (so they are adjacent) or $2$.
	\end{itemize}
\end{itemize}

We also define \emph{cable} given three distinct pairwise non-adjacent vertices $x$, $y$, $z$ in $G$. An \emph{$(x,y,z)$-cable} in $G$ is an induced subgraph $H$ of $G$ such that:
\begin{itemize}
	\item $H$ is a path from $x'$ to $z'$ going through $y'$ for some permutation $(x',y',z')$ of $\{x,y,z\}$.
	\item For any vertex $v$ in $G\setminus H$:
	\begin{itemize}
		\item $v$ has no neighbor in $H$ or exactly one neighbor in $H$, or
		\item $v$ has exactly two neighbors $v_1$, $v_2$ in $H$ such that for some $t\in \{x',z'\}$, $v_1$, $v_2$ are in the path $y'Ht$ and their distance in $H$ is $1$ or $2$, or
		\item $v$ has exactly three neighbors in $H$, which are $y'$ and two neighbors of $y'$ in $H$.
	\end{itemize}
\end{itemize}

Note that the existence of an $(x,y,z)$-antenna or $(x,y,z)$-cable in $G$ implies that there is no vertex in $G$ adjacent to all three vertices $x$, $y$, $z$. We will use the following algorithm, which is a direct consequence of Steiner problem in graphs for fixed number of terminals: 

\begin{noname} \label{minimum connected}
There is an algorithm with the following specifications:
\begin{itemize}
\item \emph{\bf Input}: A graph $G$, a subset $X\subseteq V(G)$ of size $k$ $(k$ is fixed$)$.
\item \emph{\bf Output}: A minimum subgraph of $G$ connecting every vertex in $X$ $($minimum with respect to number of vertices$)$.
\item \emph{\bf Running time}: $O(n^3)$.
\end{itemize}
\end{noname}

\begin{proof}
By considering $X$ as the set of terminals and the weight of every edge is $1$, the solution for Steiner problem in $G$ with $k$ terminals gives a tree $T$ (a subgraph of $G$) connecting $X$ with minimum number of edges. Since $T$ is a tree, the number of its vertices differs exactly one from the number of its edges, therefore graph $G$ induced by $V(T)$ is also the solution for \ref{minimum connected}. An $O(n^3)$ algorithm for Steiner problem in graphs with fixed number of terminals is given in \cite{DW71}.
\end{proof}

\begin{noname} \label{tree or little more}
Given a connected graph $G$ and three vertices $x,y,z\in V(G)$, a minimum subgraph $H$ of $G$ connecting $x,y,z$ induces either:
\begin{enumerate}
	\item A path, or 
	\item A tree containing exactly one claw, or
	\item The line graph of a tree containing exactly one claw.
\end{enumerate}  
\end{noname}

\begin{proof}
If there are more edges, we would find a smaller subgraph in $G$ connecting $x,y,z$, contradiction.
\end{proof}

From now on, we always denote by $G,u,x,y,z$ the input of Algorithm \ref{detect radar} and denote by $G'$ the graph $G\setminus (N[u]\setminus \{x,y,z\})$. The following algorithm shows that we can detect some nice structures in $G'$ in polynomial time.

\begin{noname} \label{antenna or cable}
There is an algorithm with the following specifications:
\begin{itemize}
\item \emph{\bf Input}: $G,u,x,y,z$.
\item \emph{\bf Output}: One of the followings:
	\begin{itemize}
		\item An ISK4 in $G$, or
		\item Conclude that there is no $(x,y,z)$-radar in $G'$, or
		\item A vertex $v\in G'$ adjacent to all three vertices $x,y,z$, or
		\item An $(x,y,z)$-antenna $H$ in $G'$, or
		\item An $(x,y,z)$-cable $H$ in $G'$.
	\end{itemize}
\item \emph{\bf Running time}: $O(n^3)$.
\end{itemize}
\end{noname}

\begin{proof}
First, we check if $x,y,z$ are connected in $G'$ in $O(n^2)$. If they are not connected, conclude that there is no $(x,y,z)$-radar in $G'$. Now suppose that they are connected, we can find a minimum induced subgraph $H$ of $G$ connecting $x,y,z$ by  Algorithm \ref{minimum connected}. By \ref{tree or little more}, if $H$ does not induce a path or a tree, output $H\cup \{u\}$ as an ISK4 in $G$. Therefore, we may assume that $H$ induces a path or a tree. If $H$ contains a vertex adjacent to both $x,y,z$, output that vertex and stop. Otherwise, we will prove that $H$ must be an $(x,y,z)$-antenna or an $(x,y,z)$-cable in $G'$, or $G$ contains an ISK4. It is clear that now $H$ must have the same induced structure as an antenna or a cable. We are left to prove that the attachment of a vertex $v\in G'\setminus H$ also satisfies the conditions in both cases:
\begin{itemize}[leftmargin=*]
	\item Case $1$: $H$ has the same induced structure as an $(x,y,z)$-antenna. Let $c$ be the center of the only claw in $H$. Let $x',y',z'$ be three neighbors of $c$ such that $x'$ ($y',z'$) is the one closest to $x$ ($y,z$, respectively) in $H$. Denote by $P_x$, $P_y$, $P_z$ the paths from $x$ to $x'$, $y$ to $y'$, $z$ to $z'$ in $H$, respectively. Let $v\in G'\setminus H$. The following is true:
	\begin{itemize}
		\item $v$ cannot have neighbors in both $P_x$, $P_y$, $P_z$.
				
		If $v$ does, $N_H(v)=\{x',y',z',c\}$ or $N_H(v)=\{x',y',z'\}$, otherwise $(H\setminus \{c,t\})\cup\{v\}$ is a graph connecting $x,y,z$ which is smaller than $H$, where $t$ is one of $\{x',y',z'\}$, contradiction. If $N_H(v)=\{x',y',z',c\}$, $\{u,v,c\}\cup P_x\cup P_y$ induces an ISK4 in $G$. If $N_H(v)=\{x',y',z'\}$, suppose that $z'\neq z$ (since $v$ is not adjacent to both $x$, $y$, $z$), then $\{u,v,c,z'\}\cup P_x\cup P_y$ induces an ISK4 in $G$.
		\item $v$ has at most two neighbors in $P_x\cup \{c\}$ (this holds for $P_y$, $P_z$ also).
		
		If $v$ has at least four neighbors in $P_x\cup \{c\}$, let $P$ be a shortest path from $x$ to $c$ in $H\cup \{v\}$, then $P\cup P_y\cup P_z$ induces a graph connecting $x$, $y$, $z$ which is smaller than $H$, contradiction. If $v$ has exactly three neighbors in $P_x\cup \{c\}$, suppose that $v$ has no neighbor in $P_z$ (since $v$ cannot have neighbors in both $P_x$, $P_y$, $P_z$), then $\{u,v,c\}\cup P_x\cup P_z$ induces an ISK4 in $G$. 
		\item $v$ cannot have neighbors in both two paths among $P_x$, $P_y$, $P_z$.
		
		W.l.o.g, suppose $v$ has neighbors in both $P_x$ and $P_y$, we might assume that $v$ has no neighbor in $P_z$. If $v$ has two neighbors in one of $P_x\cup\{c\}$ and $P_y\cup\{c\}$, suppose that is $P_x\cup \{c\}$, let $t$ be the neighbor of $v$ in $P_y$ which is closest to $c$. In this case, $\{u,v\}\cup P_x\cup P_z\cup tP_yy'$ induces an ISK4 in $G$. Therefore, $v$ has exactly one neighbor in $P_x$ and one neighbor in $P_y$ and $H\cup \{u,v\}$ induces an ISK4 in $G$.
		
		\item If $v$ has exactly two neighbors in $P_x\cup \{c\}$, they must be of distance $1$ or $2$ in $H$.
		
		Otherwise, we find a graph connecting $x$, $y$, $z$ smaller than $H$, contradiction. 
	\end{itemize}
	
	\item Case $2$: $H$ has the same induced structure as an $(x,y,z)$-cable. Suppose that $H$ is a path from $x$ to $z$ going through $y$. Let $x'$, $z'$ be the two neighbors of $y$ in $H$ such that $x'$ is closer to $x$ in $H$. Denote by $P_x$, $P_z$ the paths from $x$ to $x'$, $z$ to $z'$ in $H$, respectively. Let $v\in G'\setminus H$. The following is true:
	\begin{itemize}
		\item $v$ has at most two neighbors in $P_x\cup \{y\}$.
		
		If $v$ has four neighbors in $P_x\cup \{y\}$, let $P$ be a shortest path from $x$ to $y$ in $H\cup \{v\}$, then $P\cup P_z$ is a subgraph connecting $x,y,z$ which is smaller than $H$, contradiction. If $v$ has exactly three neighbors in $P_x\cup \{y\}$, then $\{u,v,y\}\cup P_x$ induces an ISK4 in $G$.
		\item If $v$ has neighbors in both $P_x$, $P_z$, then $N_H(v)=\{x',y,z'\}$.
		
		We first show that $v$ is adjacent to $y$. Suppose that $v$ is not adjacent to $y$. If $v$ has two neighbors in $P_x$, let $t$ be the neighbor of $v$ in $P_z$ which is closest to $y$. In this case, $\{u,v,y\}\cup P_x\cup tP_zz'$ induces an ISK4 in $G$. Therefore, $v$ has exactly one neighbor in $P_x$ and one neighbor in $P_y$. But now, $\{u,v\}\cup H$ induces an ISK4 in $G$.
		
		Now, $v$ is adjacent to $y$. Since $v$ has at most two neighbors in $P_x\cup \{y\}$ and two neighbors in $P_z\cup \{y\}$, $v$ has exactly one neighbor in $P_x$ and one neighbor in $P_z$. If $v$ is not adjacent to $x'$, let $t$ be the neighbor of $v$ in $P_x$. Now $\{u,v,y\}\cup P_z\cup xP_xt$ induces an ISK4 in $G$. Therefore, $N_H(v)=\{x',y,z'\}$.
		\item If $v$ has exactly two neighbors in $P_x$, they must be of distance $1$ or $2$ in $H$.
		
		Otherwise, we find a graph connecting $x$, $y$, $z$ smaller than $H$, contradiction.
	\end{itemize}
\end{itemize}
\end{proof}

Actually, there is an alternative way to implement Algorithm \ref{antenna or cable} more efficiently by not using \ref{minimum connected}. Basically, we only have to consider a shortest path $P_{xy}$ from $x$ to $y$, then find a shortest path from $z$ to $P_{xy}$. By that we would obtain immediately an $(x,y,z)$-antenna or $(x,y,z)$-cable. However, we use \ref{minimum connected} since it gives us a more convenient proof. The first case we need to handle in Algorithm \ref{antenna or cable} is when there is some vertex $v$ adjacent to both $x$, $y$ and $z$.

\begin{noname} \label{handle three adjacent}
There is an algorithm with the following specifications:
\begin{itemize}
\item \emph{\bf Input}: $G,u,x,y,z$, some vertex $v\in G'$ adjacent to $x$, $y$, $z$.
\item \emph{\bf Output}: One of the followings:
	\begin{itemize}
		\item An ISK4 in $G$, or
		\item Conclude that $v$ is not contained in any $(x,y,z)$-radar in $G'$.
	\end{itemize}
\item \emph{\bf Running time}: $O(n^2)$.
\end{itemize}
\end{noname}

\begin{proof}
It is not hard to see the following: $v$ is contained in some $(x,y,z)$-radar in $G'$ iff there exists a path from $y$ to $z$ in $G_x=G'\setminus ((N[x]\cup N[v])\setminus \{y,z\})$ (up to a relabeling of $x$, $y$, $z$). Therefore, we only have to test if $y$ and $z$ are connected in $G_x$ (and symmetries). If we find some path $P$ from $y$ to $z$ in $G_x$, output $\{u,x,v\}\cup P$ as an ISK4. If no such path exists, we can conclude that $v$ is not contained in any $(x,y,z)$-radar in $G'$. Since we only have to test the connection three times (between $y$ and $z$ in $G_x$, and symmetries), the running time of this algorithm is $O(n^2)$. 
\end{proof}

We also have the following algorithm to handle with antenna.

\begin{noname} \label{handle antenna}
There is an algorithm with the following specifications:
\begin{itemize}
\item \emph{\bf Input}: $G,u,x,y,z$, an $(x,y,z)$-antenna $H$ in $G'$.
\item \emph{\bf Output}: One of the followings:
	\begin{itemize}
		\item An ISK4 in $G$, or
		\item Conclude that there is no $(x,y,z)$-radar in $G'$, or
		\item Some vertex $c\in G'$ which is not contained in any $(x,y,z)$-radar in $G'$.
	\end{itemize}
\item \emph{\bf Running time}: $O(n^4)$.
\end{itemize}
\end{noname}

\begin{proof}
Denote by $c$, $x'$, $y'$, $z'$, $P_x$, $P_y$, $P_z$ the elements of $H$ as in the definition of an antenna. In this proof, we always denote by $N(X)$ the neighbor of $X$ in $G'$. First, we prove that any path connecting any pair of $\{x,y,z\}$ in $G'\setminus c$  which contains at most two neighbors of $c$ certifies the existence of an ISK4 in $G$. Such a path can be found  by generating every pair $(v,t)$ of neighbors of $c$ in $G'$, and for each pair, find a shortest path between each pair of $\{x,y,z\}$ in $G'\setminus (N[c]\setminus \{v,t\})$. It is clear that if such a path is found by this algorithm, then it has at most two neighbors of $c$ and if  no path is reported, we can conclude that it does not exist. Since we have $O(n^2)$ pairs $(v,t)$ and finding a shortest path between some pair of vertices in a graph takes $O(n^2)$, this algorithm runs in $O(n^4)$. Now we prove that such a path certifies the existence of an ISK4. Let $P$ be a path between some pair in $\{x,y,z\}$ that contains at most two neighbors of $c$, w.l.o.g assume that $P$ is from $x$ to $y$. We say that a path $Q$ is a \emph{$(P_x,P_y)$-connection} if one end of $Q$ is in $N(P_x)$, the other end is in $N(P_y)$ and $Q^*\cap (N(P_x)\cup N(P_y))=\emptyset$ (we make  symmetric definitions for $(x,z)$ and $(y,z)$). We also say that a path $Q$ is \emph{$S$-independent} for some $S\subseteq V(G')$ if $Q\cap N[S]=\emptyset$. We consider following cases:
\begin{enumerate}
	\item $P$ contains no neighbor of $c$.
	 
	It is clear that there exists a subpath $P'$ of $P$ such that $P'$ is a $(P_x,P_y)$-connection. Furthermore, we may assume that $P'$ is $P_z$-independent since otherwise there exists some subpath $P''$ of $P'$ which is a $(P_x,P_z)$-connection and is $P_y$-independent. Let $x''$, $y''$ be two ends of $P'$ which are in $N(P_x)$ and $N(P_y)$, respectively. In this case $x''$ and $y''$ are not adjacent to $c$ since $P'$ contains no neighbor of $c$. We have the following cases based on the attachment on an antenna:
	\begin{enumerate}
		\item $x''$ and $y''$, each has exactly one neighbor in $P_x$ and $P_y$, respectively. Then $\{u\}\cup P'\cup H$ induces an ISK4 in $G$.
		\item $x''$ has exactly one neighbor in $P_x$ and $y''$ has exactly two neighbors in $P_y$ (or symmetric). Then $\{u,c\}\cup P'\cup P_x\cup P_y$ induces an ISK4 in $G$.
		\item $x''$ and $y''$, each has exactly two neighbors in $P_x$ and $P_y$, respectively. Let $t$ be the neighbor of $x$ in $P_x$ which is closer to $c$. Then $\{u,c\}\cup P'\cup tP_xx'\cup P_y\cup P_z$ induces an ISK4 in $G$.
	\end{enumerate}	 
	 
	\item $P$ contains exactly one neighbor of $c$.
	
	Similar to the argument of the previous case, there exists a path $P'$ with two ends $x''$ and $y''$ such that $P'$ is a $(P_x,P_y)$-connection and is $P_z$-independent. We may assume that $c$ has exactly one neighbor $c'$ in $P'$, otherwise we are back to previous case. In this case, at most one vertex in $\{x'',y''\}$ can be adjacent to $c$ (in other words, at most one vertex in $\{x'',y''\}$ can be identical to $c'$). We consider the following cases:
	\begin{enumerate}
		\item Each of $\{x'',y''\}$ has exactly one neighbor in $P_x\cup \{c\}$, and therefore exactly one neighbor in $P_x$. Then $\{u,c\}\cup P'\cup P_x\cup P_y$ induces an ISK4 in $G$.
		\item $x''$ has exactly one neighbor in $P_x\cup \{c\}$ (this neighbor must be in $P_x$) and $y''$ has exactly two neighbors in $P_x\cup \{c\}$ (or symmetric). If $y''$ is adjacent to $c$, then $\{u,c\}\cup P'\cup P_x\cup P_y$ induces an ISK4 in $G$. Otherwise $y''$ has two neighbors in $P_y$ and $\{u,c\}\cup c'P'y''\cup P_x\cup P_y$ induces an ISK4 in $G$.
		\item Each of $\{x'',y''\}$ has exactly two neighbors in $P_x\cup \{c\}$. Since at most one of them is adjacent to $c$, we might assume that $y$ is not adjacent to $c$. Then $\{u,c\}\cup c'P'y''\cup P_y\cup P_z$ induces an ISK4 in $G$. 
	\end{enumerate}		
	
	\item $P$ contains exactly two neighbors of $c$.
	
	We may assume that $P$ is $P_z$-independent since otherwise we have some subpath of $P$ which is a $(P_z,P_x)$-connection or $(P_z,P_y)$-connection and contains at most one neighbor of $c$ that we can argue like previous cases. Therefore, $\{u,c\}\cup P \cup P_z$ induces an ISK4.	
\end{enumerate}

It is easy to see that above argument can be turned into an algorithm to output an ISK4 in each case. Now we can describe our algorithm for \ref{handle antenna}. First, test if there exists a path in $G'\setminus c$ between some pair of $\{x,y,z\}$ which contains at most two neighbors of $c$:
\begin{enumerate}
	\item \label{antenna:case 1} If such a path exists, output the corresponding ISK4 in $G$.
	\item \label{antenna:case 2} If no such path exists, test the connection between each pair of $\{x,y,z\}$ in $G'\setminus c$:
	\begin{enumerate}
		\item \label{antenna:case 2a} If $\{c\}$ is a cutset in $G'$ disconnecting some pair of $\{x,y,z\}$, then conclude that  there is no $(x,y,z)$-radar in $G'$.
		\item \label{antenna:case 2b} Otherwise, conclude that $c$ is the vertex not contained in any $(x,y,z)$-radar in $G'$.
	\end{enumerate}
\end{enumerate}
Now we explain why this algorithm is correct. If Case \ref{antenna:case 1} happens, it outputs correctly the ISK4 by the argument above. If Case \ref{antenna:case 2} happens, we know that there are only two possible cases for  the connection between each pair of $\{x,y,z\}$ in $G'\setminus c$, for example for $(x,y)$:
\begin{itemize}
	\item $x$ and $y$ are not connected in $G'\setminus c$, or
	\item Every path from $x$ to $y$ in $G'\setminus c$ contains at least three neighbors of $c$.
\end{itemize}
Therefore, Case \ref{antenna:case 2a} corresponds to one of the following cases, both lead to the conclusion that there is no $(x,y,z)$-radar in $G'$:
\begin{itemize}
	\item Each pair of $\{x,y,z\}$ is not connected in $G'\setminus c$.
	\item $x$ is not connected to $\{y,z\}$, while $y$ and $z$ are still connected in $G'\setminus c$ (or symmetric). In this case every path from $y$ to $z$ in $G'\setminus c$ contains at least three neighbors of $c$.
\end{itemize}
If Case \ref{antenna:case 2b} happens, we know that each pair of $\{x,y,z\}$ is still connected in $G'\setminus c$ and furthermore every path between them contains at least three neighbors of $c$. This implies that $c$ is not contained in any $(x,y,z)$-radar, since if $c$ is in some $(x,y,z)$-radar, we can easily find a path between some pair of $\{x,y,z\}$ in that radar containing at most two neighbors of $c$, contradiction.

The complexity of the whole algorithm is still $O(n^4)$ since we can find an ISK4 in Case \ref{antenna:case 1} in $O(n^2)$ and test the connection in Case \ref{antenna:case 2} in $O(n^2)$.  
\end{proof}

The next algorithm deals with cable.

\begin{noname} \label{handle cable}
There is an algorithm with the following specifications:
\begin{itemize}
\item \emph{\bf Input}: $G,u,x,y,z$, an $(x,y,z)$-cable $H$ in $G'$.
\item \emph{\bf Output}: One of the followings:
	\begin{itemize}
		\item An ISK4 in $G$, or
		\item Conclude that there is no $(x,y,z)$-radar in $G'$, or
		\item Some vertex $c\in G'$ which is not contained in any $(x,y,z)$-radar in $G'$.
	\end{itemize}
\item \emph{\bf Running time}: $O(n^4)$.
\end{itemize}
\end{noname} 

\begin{proof}
W.l.o.g we can assume that cable $H$ is a path from $x$ to $z$ containing $y$. Let $x'$ be the neighbor of $y$ in $H$ which is closer to $x$ and $z'$ be the other neighbor of $y$ in $H$. Let $P_x=xHx'$ and $P_z=zHz'$. In this proof, we denote by $N(X)$ the neighbor of $X$ in $G'$. We also say that a path $Q$ is a \emph{$(P_x,P_z)$-connection} if one end of $Q$ is in $N(P_x)$, the other end is in $N(P_z)$ and $Q^*\cap (N(P_x)\cup N(P_z))=\emptyset$. Before the algorithm, we first prove the followings:

\begin{enumerate}[(1),leftmargin=*]
	\item \label{no neighbor} Every path $P$ from $x$ to $z$ in $G'\setminus y$ containing no neighbor of $y$ certifies an ISK4 in $G$.
	
	Let $P'$ be a subpath of $P$ such that $P'$ is a $(P_x,P_z)$-connection. Let $x''$ and $z''$ be two ends of $P'$ such that $x''\in N(P_x)$ and $z''\in N(P_z)$. Since $P'$ has no neighbor of $y$, both $x''$ and $z''$ are not adjacent to $y$. We consider the following cases based on the attachment on a cable:
	\begin{enumerate}
		\item $x''$ and $z''$, each has exactly one neighbor in $P_x$ and $P_z$, respectively. Then $\{u\}\cup H\cup P'$ induces an ISK4 in $G$.
		\item $x''$ has exactly two neighbors in $P_x$ and $z''$ has exactly one neighbor in $P_z$ (or symmetric). Let $t$ be the neighbor of $z''$ in $P_z$.
		\begin{itemize}
			\item If $t\neq z$ then $\{u,y\}\cup P'\cup P_x\cup tP_zz'$ induces an ISK4 in $G$.
			\item If $t=z$ then $\{u,y,z\}\cup P'\cup P_x$ induces an ISK4 in $G$.
		\end{itemize}
		\item $x''$ and $z''$, each has exactly two neighbors in $P_x$ and $P_z$, respectively. Let $t$ be one of the two neighbors of $z''$ which is closer to $y$. Then $\{u,y\}\cup P'\cup P_x\cup tP_zz'$ induces an ISK4 in $G$. 
	\end{enumerate}

	\item \label{two neighbors} Every path $P$ from $x$ to $z$ in $G'\setminus y$ containing exactly two neighbors of $y$ certifies an ISK4 in $G$.
	
	It is clear since $\{u,y\}\cup P$ induces an ISK4 in $G$.

	\item \label{one neighbor} Assume that every path from $x$ to $z$ in $G'\setminus y$ contains at least one neighbor of $y$. If there exists some path from $x$ to $z$ in $G'\setminus y$ containing exactly one neighbor of $y$, then a shortest such path $P$ satisfies that $P\cup\{y\}$ is an $(x,y,z)$-antenna in $G'$, or $G$ contains an ISK4.
	
	It is clear that $P\cup\{y\}$ has the same induced structure as an antenna, we only have to prove the attachment on it. Let $c$ be the only neighbor of $y$ on $P$ and $x'$, $z'$ be the two neighbors of $c$ different from $y$ such that $x'$ is the one closer to $x$ in $P$. Denote $P_x=xPx'$, $P_z=zPz'$. Let $v$ be a vertex in $G'\setminus (P\cup\{y\})$, we consider the following cases:
	\begin{itemize}
		\item $v$ is not adjacent to $y$. The following is true:
		\begin{itemize}
			\item $v$ cannot have neighbors on both $P_x$ and $P_z$.
			
			If $v$ does, there exists a path in $G'$ from $x$ to $z$ (passing through $v$) containing no neighbor of $y$, contradiction.
			\item $v$ has at most two neighbors in $P_x\cup \{c\}$.
			
			If $v$ has at least three neighbors in $P_x\cup \{c\}$, they must be exactly three consecutive neighbors in $P$, otherwise there exists a shorter path than $P$ satisfying the assumption. But if $v$ has three consecutive neighbors in $P_x\cup \{c\}$, then $\{u,v\}\cup P$ induces an ISK4.
			\item If $v$ has exactly two neighbors in $P_x\cup \{c\}$, they must be of distance $1$ or $2$ in $P$.
			
			Otherwise, we have a shorter path than $P$ (passing through $v$) satisfying the assumption. 
		\end{itemize}
		\item $v$ is adjacent to $y$. The following is true:
		\begin{itemize}
			\item $v$ cannot have neighbors on both $P_x$ and $P_z$.
			
			If $v$ does, $N(v)\cap P_x=\{x'\}$ and $N(v)\cap P_z=\{z'\}$, otherwise there exists a shorter path than $P$ (passing through $v$) satisfying the assumption. If $v$ is adjacent to $c$, then $\{u,v\}\cup P$ induces an ISK4. If $v$ is not adjacent to $c$, since $v$ cannot adjacent to both $x$ and $z$ (by definition of cable), assume $v$ is not adjacent to $x$ (or equivalently $x\neq x'$). In this case, $\{u,v,y,x'\}\cup P_z$ induces an ISK4 in $G$.
			\item $v$ cannot have at least three neighbors in $P_x\cup \{c\}$.
			
			If $v$ does, there exists a path (passing through $v$) from $x$ to $z$ containing exactly two neighbors of $y$ (which are $v$ and $c$). This path certifies an ISK4 by \ref{two neighbors}.
			
			\item $v$ cannot have exactly two neighbors in $P_x\cup \{c\}$.
			
			If $v$ does, $\{u,v,y,c\}\cup P_x$ induces an ISK4.
			\item If $v$ has exactly one neighbor in $P_x\cup \{c\}$, it must be $c$.
			
			If $v$ has exactly one neighbor in $P_x\cup \{c\}$ which is not $c$, then $\{u,v,y\}\cup P$ induces an ISK4.
		\end{itemize}
	\end{itemize}
	The above discussion shows that either $G$ contains an ISK4 (and we can detect in $O(n^2)$), or $P\cup \{y\}$ is an $(x,y,z)$-antenna in $G'$.
\end{enumerate}

Now we describe our algorithm for \ref{handle cable}:
\begin{enumerate}
	\item \label{step 0}  Test if there exists a path $P$ from $x$ to $z$ in $G'\setminus y$ containing no neighbor of $y$.
	\begin{enumerate}
		\item If such a path exists, output an ISK4 by the argument in \ref{no neighbor}.
		\item If no such path exists, move to the next step.
	\end{enumerate}
	\item \label{step 1} Find a shortest path $P$ from $x$ to $z$ in $G'\setminus y$ containing exactly one neighbor of $y$ if such a path exists.
	\begin{enumerate}
		\item \label{step 1a} If such a path $P$ exists, by the argument in \ref{one neighbor}, either we detect an ISK4 in $G$, output it and stop, or we find an $(x,y,z)$-antenna $P\cup\{y\}$ in $G'$, run Algorithm \ref{handle antenna} given this antenna as input, output the corresponding conclusion.    
		\item If no such path exists, move to the next step.
	\end{enumerate}
	\item \label{step 2} Test if there exists a path $P$ from $x$ to $z$ in $G'\setminus y$ containing exactly two neighbors of $y$.
	\begin{enumerate}
		\item If such a path $P$ exists, output an ISK4 in $G$ by the argument in \ref{two neighbors}.
		\item If no such path exists, conclude there is no $(x,y,z)$-radar in $G'$ (Since at this point, every path from $x$ to $z$ in $G'\setminus y$ contains at least three neighbors of $y$).
	\end{enumerate}
\end{enumerate}
		
Step \ref{step 0} can be done in $O(n^2)$ by checking the connection between $x$ and $z$ in $G'\setminus N[y]$. Step \ref{step 1} runs in $O(n^3)$ by generating every neighbor $t$ of $y$ and for each $t$, find a shortest path between $x$ and $z$ in $G'\setminus (N[y]\setminus \{t\})$. And pick the shortest one over all such paths. Since we call the Algorithm \ref{handle antenna}, step \ref{step 1a} takes $O(n^4)$. Step \ref{step 2} can be done in $O(n^4)$	by generating every pair $(t,w)$ of neighbors of $y$ and for each pair $(t,w)$, check the connection between $x$ and $z$ in $G'\setminus (N[y]\setminus \{t,w\})$. Therefore, the total running time of Algorithm \ref{handle cable} is $O(n^4)$.
\end{proof}

\section{Proof of \ref{detect radar}} \label{S:4}

Now we sum up everything in previous section and describe the algorithm for \ref{detect radar}:

\begin{enumerate}
	\item Run Algorithm \ref{antenna or cable}. Output is one of the followings:
	\begin{enumerate}
		\item An ISK4 in $G$: output it and stop.
		\item Conclude that there is no $(x,y,z)$-radar in $G'$ and stop.
		\item A vertex $v$ adjacent to $x$, $y$, $z$: run Algorithm \ref{handle three adjacent} with $v$ as input. Output is one of the followings:
		\begin{enumerate}
			\item An ISK4 in $G$: output it and stop.
			\item Conclude that $v$ is not contained in any $(x,y,z)$-radar in $G'$: Run Algorithm \ref{detect radar} recursively for $(G\setminus v,u,x,y,z)$.
		\end{enumerate}
		\item An $(x,y,z)$-antenna $H$ in $G'$: run Algorithm \ref{handle antenna} with $H$ as input. Output is one of the followings:
		\begin{enumerate}
			\item An ISK4 in $G$: output it and stop.
			\item Conclude that there is no $(x,y,z)$-radar in $G'$ and stop.
			\item Some vertex $c\in G'$ which is not contained in any $(x,y,z)$-radar in $G'$: Run Algorithm \ref{detect radar} recursively for $(G\setminus c,u,x,y,z)$.
		\end{enumerate}
		\item An $(x,y,z)$-cable $H$ in $G'$: run Algorithm \ref{handle cable} with $H$ as input. Output is one of the followings:
		\begin{enumerate}
			\item An ISK4 in $G$: output it and stop.
			\item Conclude that there is no $(x,y,z)$-radar in $G'$ and stop.
			\item Some vertex $c\in G'$ which is not contained in any $(x,y,z)$-radar in $G'$: Run Algorithm \ref{detect radar} recursively for $(G\setminus c,u,x,y,z)$.
		\end{enumerate}
	\end{enumerate}
\end{enumerate}

The correctness of this algorithm is based on the correctness of the Algorithms \ref{antenna or cable}, \ref{handle three adjacent}, \ref{handle antenna} and \ref{handle cable}. Now we analyse its complexity. Let $f(n)$ be the complexity of this algorithm. Since we have five cases, each case takes $O(n^4)$ and at most a recursive call with the complexity $f(n-1)$. Therefore $f(n)\leq O(n^4)+f(n-1)$ and $f(n)=O(n^5)$.

\section{Conclusion}

In this paper, we give an $O(n^9)$ algorithm to detect induced subdivision of $K_4$ in a given graph. We believe that the complexity might be improved to $O(n^7)$ by first decomposing the graph by clique cutset until there is no $K_{3,3}$ (using decomposition theorem in \cite{LMT12}). Now every $($ISK4$,K_{3,3})$-free graph has a linear number of edges since it is $c$-degenerate by some constant $c$ as shown in \cite{LMT12}. Therefore, testing the connection in this graph takes only $O(n)$, instead of $O(n^2)$ as in the algorithm. Also, we only have to consider $O(n^3)$ triples of three independent vertices and test every possible center of that claw at the same time instead of generating all $O(n^4)$ claws. But we prefer to keep our algorithm as $O(n^9)$ since it is simple and does not rely on decomposition theorem. We leave the following open question as conclusion: 

\noindent \textbf{Open question.} Given a graph $H$ of maximum degree $3$, can we detect induced subdivision of $H$ in polynomial time?  

\subsection*{Acknowledgement} The author would like to thank Nicolas Trotignon for his help and useful discussion.

\bibliographystyle{abbrv}
\bibliography{ISK4}

\end{document}